%% file: main.tex
\documentclass{webofc}
\usepackage[varg]{txfonts}   % Web of Conferences font

\usepackage{graphicx} % Required for inserting images
\usepackage{booktabs}
\usepackage{tikz}
\usepackage{pgfplots}
\pgfplotsset{compat=1.14}
\usepackage{anyfontsize}
\usepackage{array}
\usepackage{bm}

\begin{document}
\title{Improved Absolute Polarization Calibrator for BICEP CMB Polarimeters}

\input{author_list_mmuniverse_2025.tex}

\abstract{%
Cosmic birefringence is a hypothesized parity violation in electromagnetism that predicts a frequency-independent polarization rotation as light propagates. 
This would rotate the light from the
Cosmic Microwave Background, producing an unexpected EB correlation. However, cosmic birefringence angle is degenerate with instrument polarization angle, and breaking this degeneracy requires an absolute polarization calibration. 
We calibrate the BICEP3 telescope (a 95GHz CMB polarimeter)
by observing a rotating polarized source (RPS) with both the telescope and a small test receiver called the In-Situ Absolute Angle Calibrator (ISAAC).
}
\maketitle

\section{Introduction}
The BICEP3 telescope is a small aperture refracting telescope operating at 95GHz, designed to measure CMB polarization at degree angular scales to search for signatures of cosmic inflation \cite{bkxv}. 
With the addition of an absolute polarization calibration, it can be used as a sensitive probe for cosmic birefringence.
%Our CMB maps are deep enough that at this point the systematic uncertainty on the detector calibration is the limiting factor.
With our ground-based small aperture telescope, we have the relatively rare ability to place sources directly in the far field of the telescope, enabling direct calibrations of the fielded instrument.
The Rotating Polarized Source (RPS) has been a valuable calibrator, but is constrained by certain systematic uncertainties. 
This paper outlines recent hardware upgrades and improved understanding of these systematics.

\subsection{Absolute polarization calibration}
We aim to make an absolute polarization calibration by observing a source in the far field that has extremely well-characterized polarization. 
Absolute polarization calibration at the level we are aiming for ($\sim0.03^\circ$) requires a very well-calibrated calibrator, verification of the calibration with a separate test receiver, and meticulous attention to RF details.
% Almost any systematic issue with the RPS could produce an effect larger that the cosmic birefringence we aim to measure.

\subsection{The Rotating Polarized Source}
The Rotating Polarized Source (RPS) is an electrically chopped quasi-thermal noise source, pictured below in figure \ref{fig:rps_diagram}. 
Within the shroud (label B), there is a 15dBi standard gain horn, as well as a mechanical jig for collimation.
The shroud is lined with RF absorber material to terminate any power reflected back from the wire grid.
The polarization of the source is established by the wire grid polarizer, which cleans up the polarization from the horn and provides a mechanical reference for the polarization direction of the apparatus. Further details are available in \cite{bkxviii} and \cite{spie_proc}.

\begin{figure}[b]
    \centering
    \sidecaption
    \includegraphics[trim={0 0 0 4cm},clip,width=0.5\linewidth]{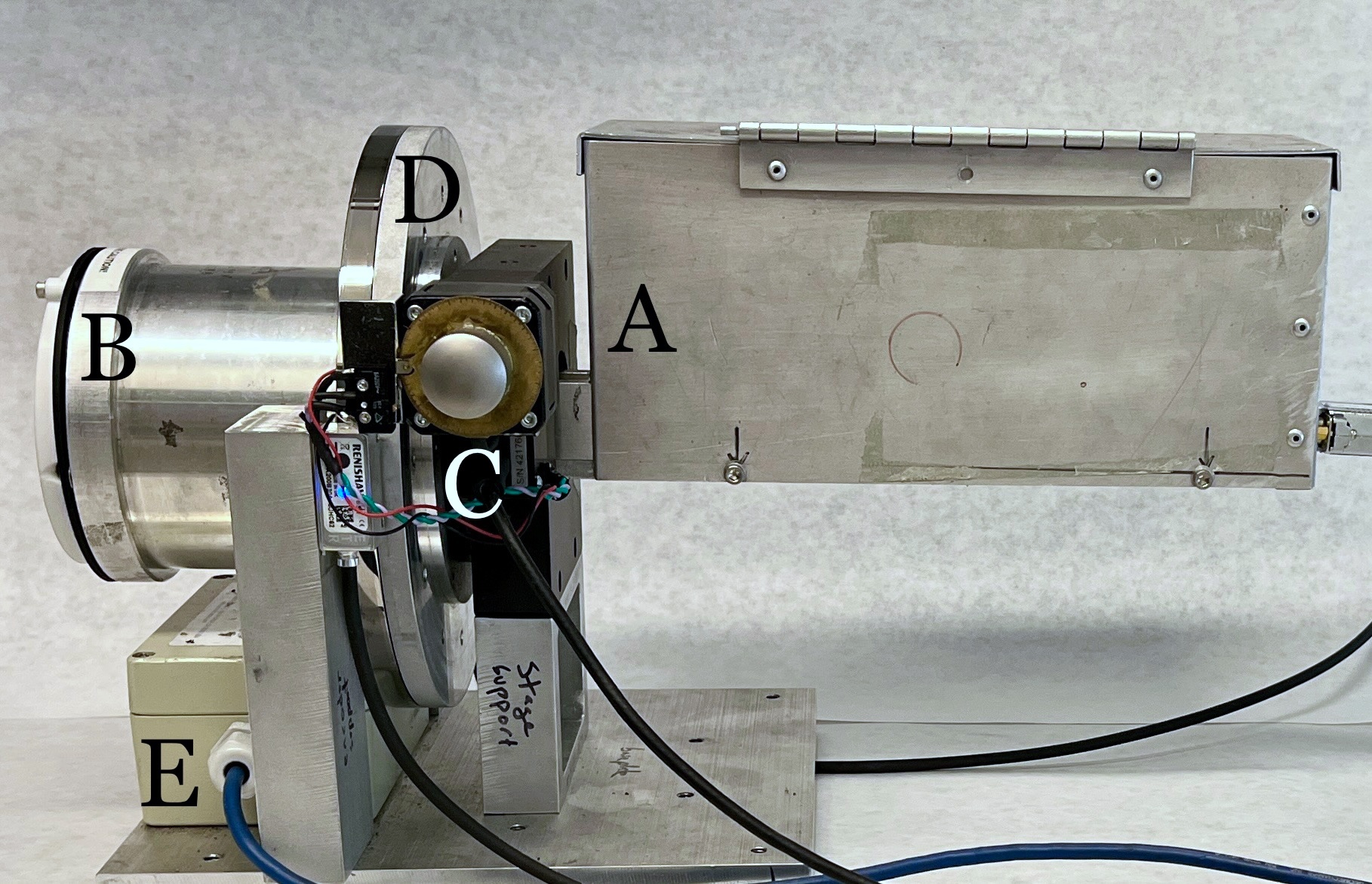}
    \caption{
    A labeled photo of the RPS.\\
    \textbf{A)} Amplified, modulated thermal noise source\\
    \textbf{B)} Wire grid polarizer (cross-pol of $-$83dB)\\
    \textbf{C)} Rotation stage\\
    \textbf{D)} Absolute rotary encoder\\
    \textbf{E)} Precision tilt-meter
    }
    \label{fig:rps_diagram}
    \vspace{-15pt}
\end{figure}

\section{Calibrating the Rotating Polarized Source}

\subsection{Benchtop calibration procedure}
The method described in this work improves on the approach in \cite{bkxviii}, which involved disassembling the stage and aligning via a Bridgeport mill. The new procedure avoids reconfiguration between calibration and deployment, simultaneously calibrates the tiltmeter and encoder, and reduces systematic error and complexity.

\begin{figure}
    \centering
    \hfill
    \includegraphics[trim={1.5cm 0.75cm 0.5cm 2cm},clip,width=0.38\linewidth]{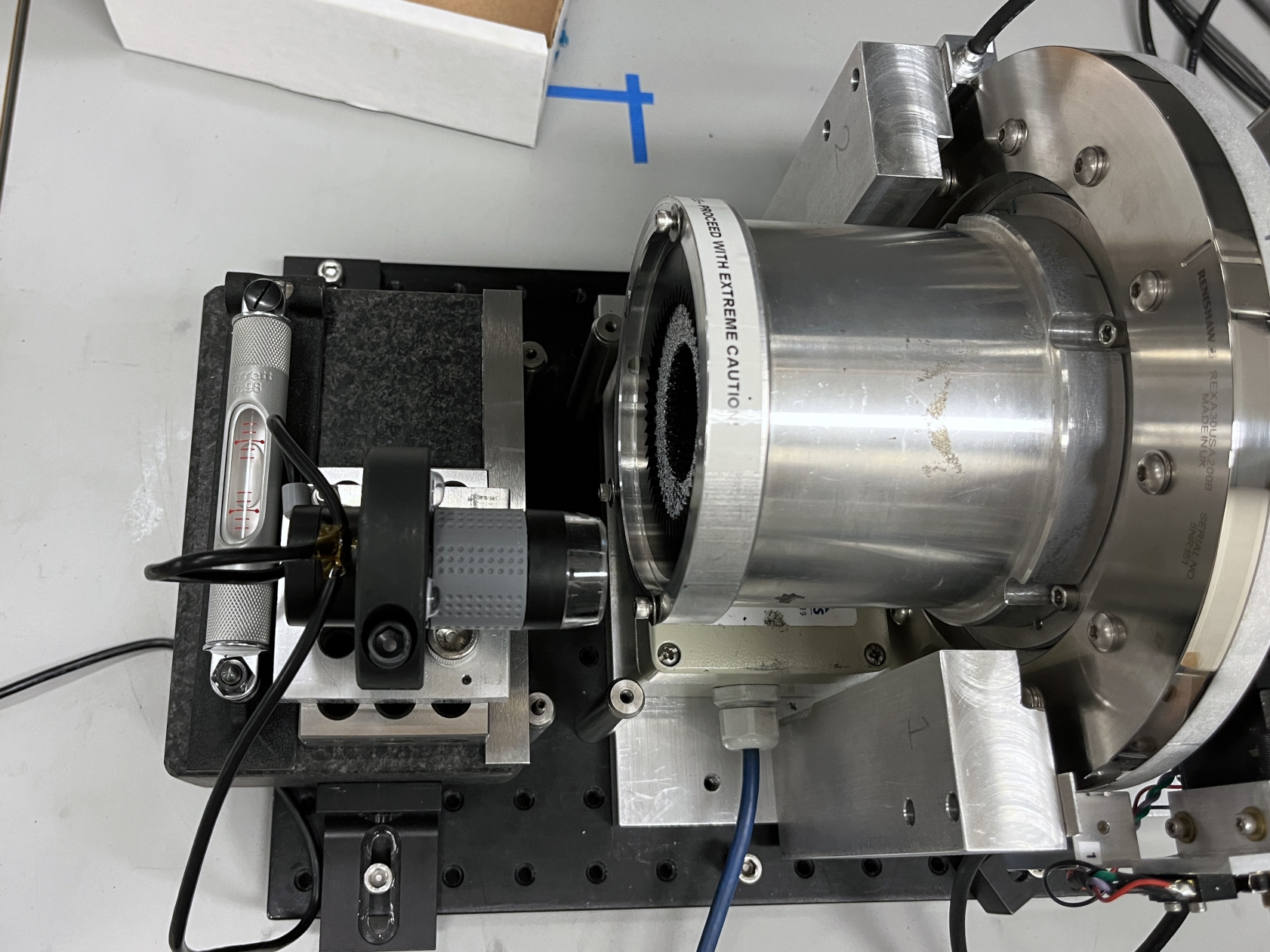}
    \hfill
    \includegraphics[trim={4cm 3cm 5cm 4.8cm},clip,width=0.35\linewidth]{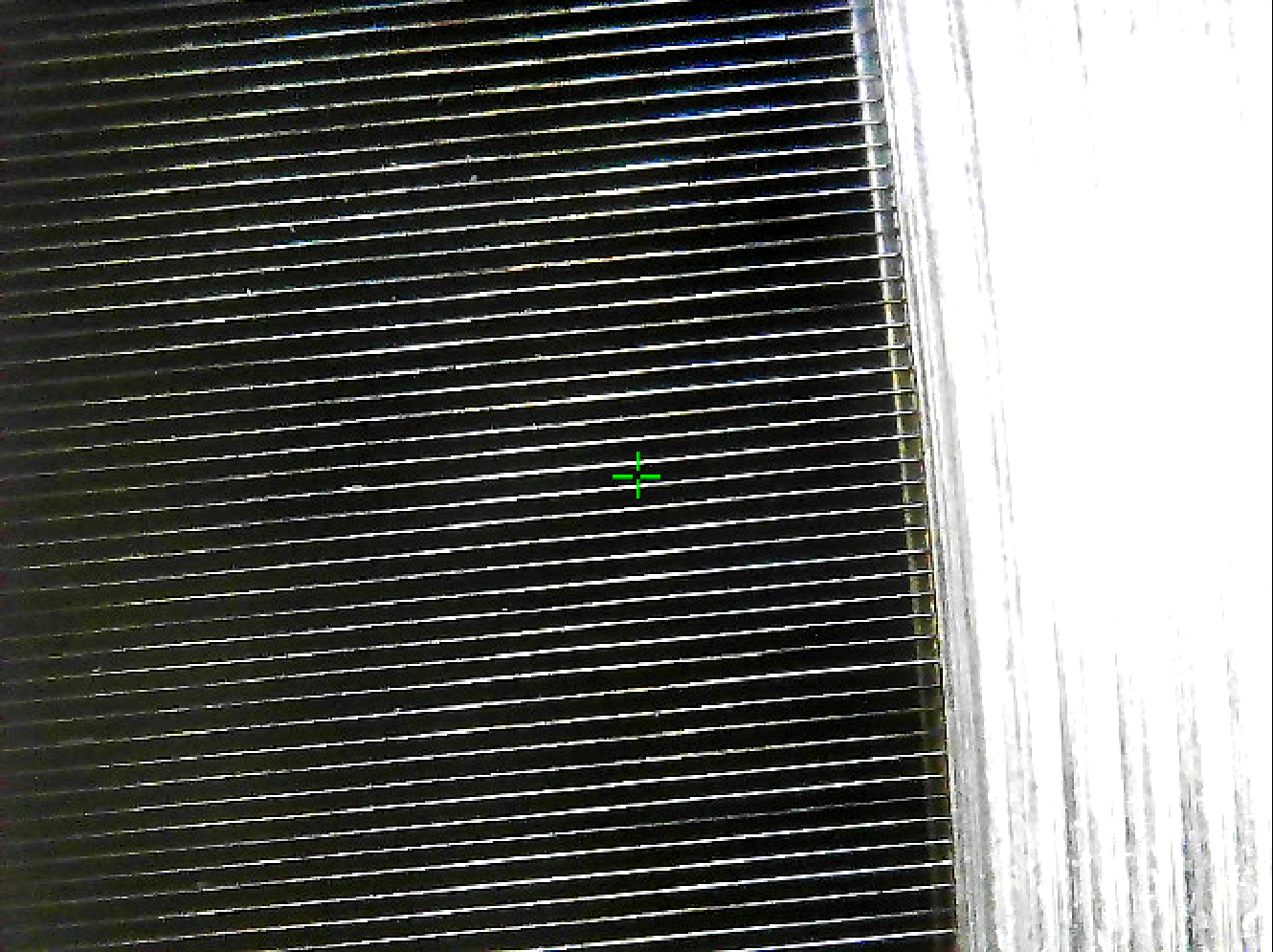}
    \hfill\hfill
    \vspace{-5pt}
    \caption{\textbf{Left}, An RPS calibration in progress. \textbf{Right}, the view through the USB microscope, where the 15$\mu$m grid wires are easily visible.}
    \label{fig:calibration}
    \vspace{-10pt}
\end{figure}

The goal of this calibration is to determine the orientation of the wire grid and thus the RPS polarization with respect to gravity. 
This calibration will ultimately be transferred to the telescope's detectors, 

The RPS and a small granite surface plate are bolted to an optical breadboard. 
A machinists level is placed on the surface plate and the breadboard is tilted until the plate is level to within 0.03$^\circ$. 
A USB microscope on a 1-2-3 block rides along the surface plate; since the plate is level, the microscope's motion is purely horizontal.
We adjust the wire grid and RPS roll until the microscope can scan the 98mm grid diameter while tracking a single 15$\mu$m wire, aligning the wire to the plate's reference surface to within 0.01$^\circ$.
Finally, we secure the wire grid into the shroud, and record the RPS tiltmeter and encoder angles.
During measurements, these calibration values, along with real-time sensor data, allow recovery of the absolute grid angle.

\subsection{Absolute optical encoder}
\begin{figure}
    \centering
    \sidecaption
    \includegraphics[width=0.45\linewidth]{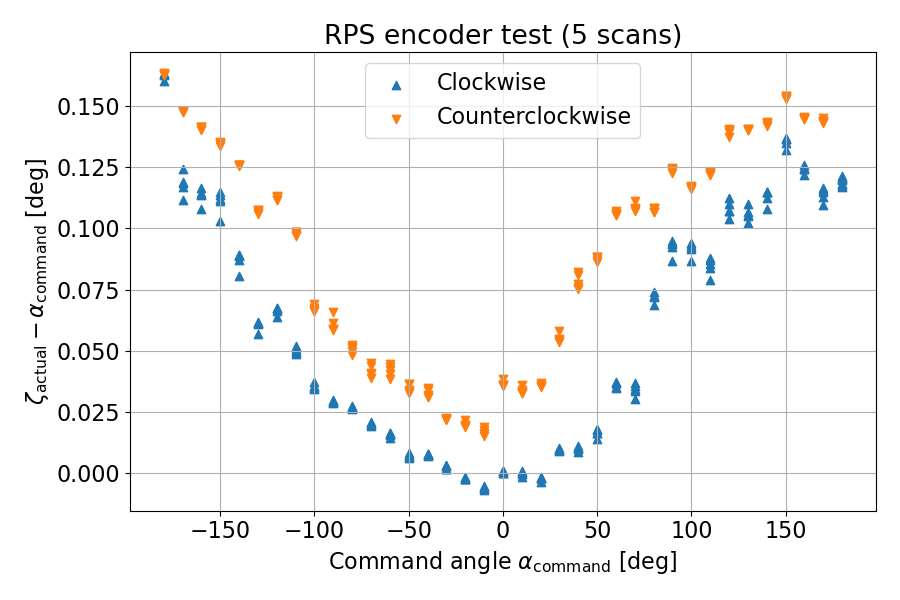}
    \caption{The absolute optical encoder reveals mechanical imperfections in the RPS rotation stage.
    This plot compares the angle commanded to the stage against the angle measured by the encoder, as it rotates through 360$^\circ$ in both directions.
    The separation between the two traces shows backlash, and the overall shape is due to elastic deformation as the uneven weight of the RPS shifts around.
    }
    \label{fig:flipflop}
\end{figure}

The addition of a $\pm$1 arcsecond absolute optical encoder breaks our dependence on the accuracy of the rotation stage by directly measuring the RPS roll angle \cite{Sjoberg2024Improved}. 
While the stepper motor is quite precise, the stepper command angle alone does not capture systematics such as backlash, which are easily measured by the encoder as shown in figure \ref{fig:flipflop}.
In \cite{bkxviii}, we attributed 0.06$^\circ$ of error to rotation stage backlash and 0.006$^\circ$ to mechanical repeatability, both of which are now replaced by the $\sim$$0.0005^\circ$ error from the encoder.

\subsection{The In-Situ Absolute Angle Calibrator}
The In-Situ Absolute Angle Calibrator, or ISAAC, is an ambient-temperature test receiver consisting of a detector diode and feed horn, along with a wire grid polarizer and tiltmeter matching the RPS \cite{bkxviii}. 
We use the ISAAC to perform an end-to-end verification of the RPS's absolute polarization calibration.
It also serves as a test receiver for in-lab measurements as we perturb and improve the RPS. 
Finally, the ISAAC includes a 35dB low noise amplifier, allowing it to co-observe the RPS with BICEP3 to provide a continuous independent verification of the RPS polarization.

\section{Calibrating BICEP3}

\begin{figure}
    \centering
    \includegraphics[width=0.8\textwidth]{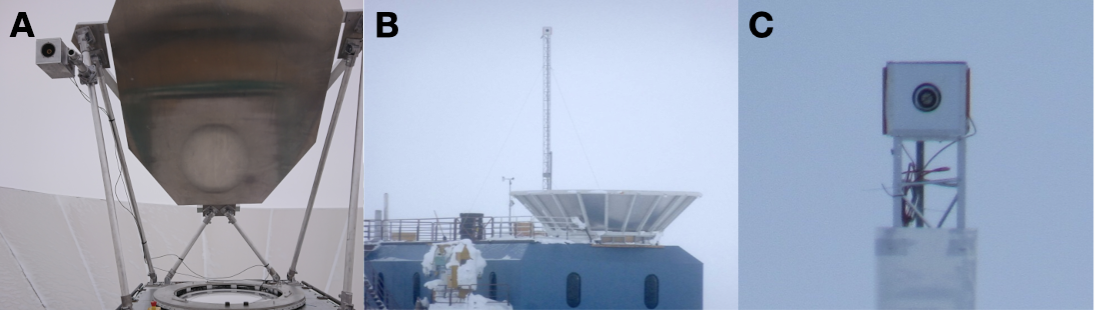}
    \vspace{-5pt}
    \caption{\textbf{A}: ISAAC mounted on the BICEP3 Far Field Flat mirror.
    \textbf{B}: RPS on MAPO mast. \textbf{C}: Detail view of RPS with alignment tabs}
    \vspace{-15pt}
\end{figure}

During a calibration campaign, the RPS is affixed to a mast 200m away, and a 45$^\circ$ mirror is installed on BICEP3 to flip its beam from zenith onto the horizon.
To transfer the calibration from the RPS to the telescope, we rotate the RPS through 13 polarization angles, rastering the telescope across it at each angle. 
For each detector, we fit a Gaussian profile to the set of 13 beams, fixing one beam center and width for the set, but allowing independent amplitudes. We then take the amplitude of each beam as the RPS amplitude at that angle.

\begin{figure}
    \centering
    \includegraphics[width=0.7\linewidth]{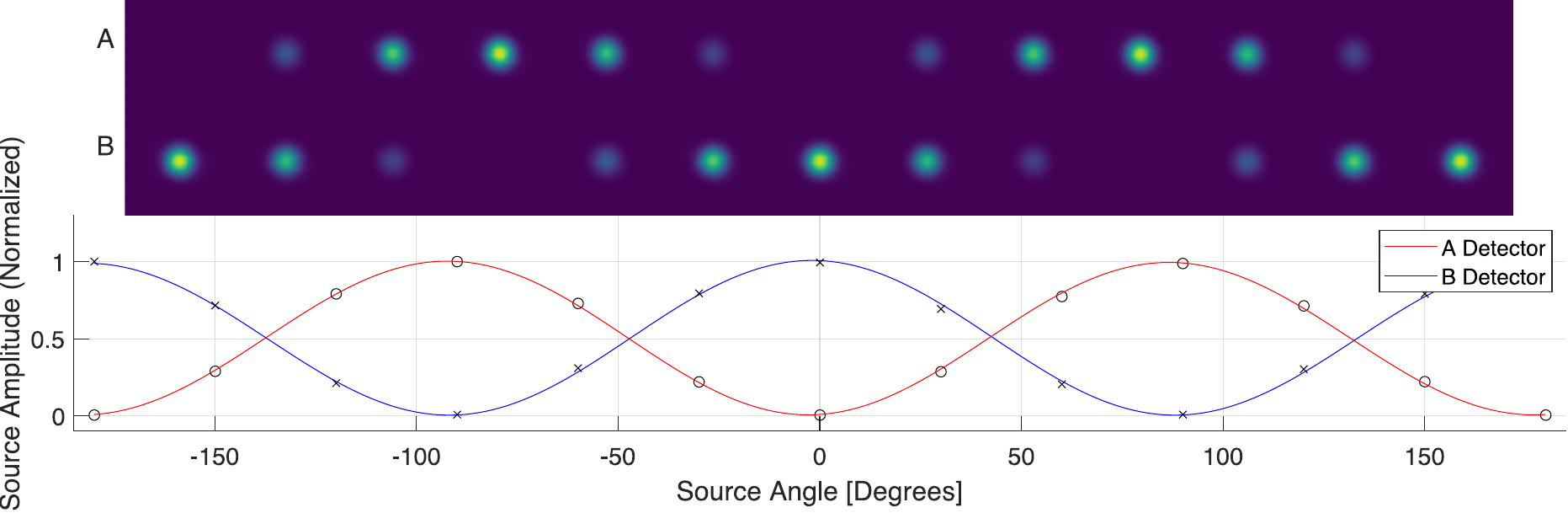}
    \vspace{-5pt}
    \caption{\textbf{Top}: Beam maps of an orthogonal pair of detectors.
    \textbf{Bottom}: Beam amplitudes (points) and modcurve model fits (lines). Figure taken from \cite{spie_proc}.}
    \vspace{-15pt}
\end{figure}

We fit each modulation curve using the following model :
\[A = {A_0}\left[\cos[2(\zeta+{\psi})]-\frac{{\epsilon}+1}{{\epsilon}-1}\right] (n_1\cos{\zeta} + n_2\sin{\zeta}+1) \]
where A is the amplitude of the curve, and depends on the following parameters:
$\bm{\zeta}$ RPS grid angle with respect to gravity;
$\bm{A_0}$ Overall modulation curve amplitude;
$\bm{\epsilon}$ Detector cross-polar response;
$\bm{n_1}, \bm{n_2}$ Nuisance parameters (source miscollimation);
$\bm{\psi}$ Angle between source and detector co-pol axes.
For more details on this procedure, see \cite{bkxviii}.

\subsection{Successful measurements of BICEP3}
\begin{figure}
    \centering
    \includegraphics[width=0.4\linewidth]{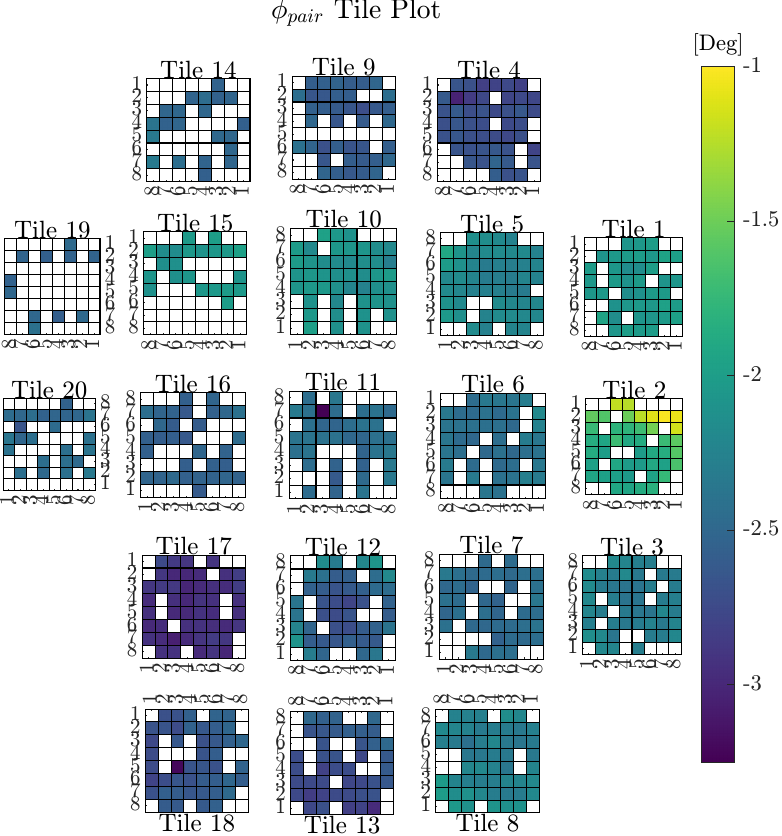}
    \hfill
    \includegraphics[width=0.4\linewidth]{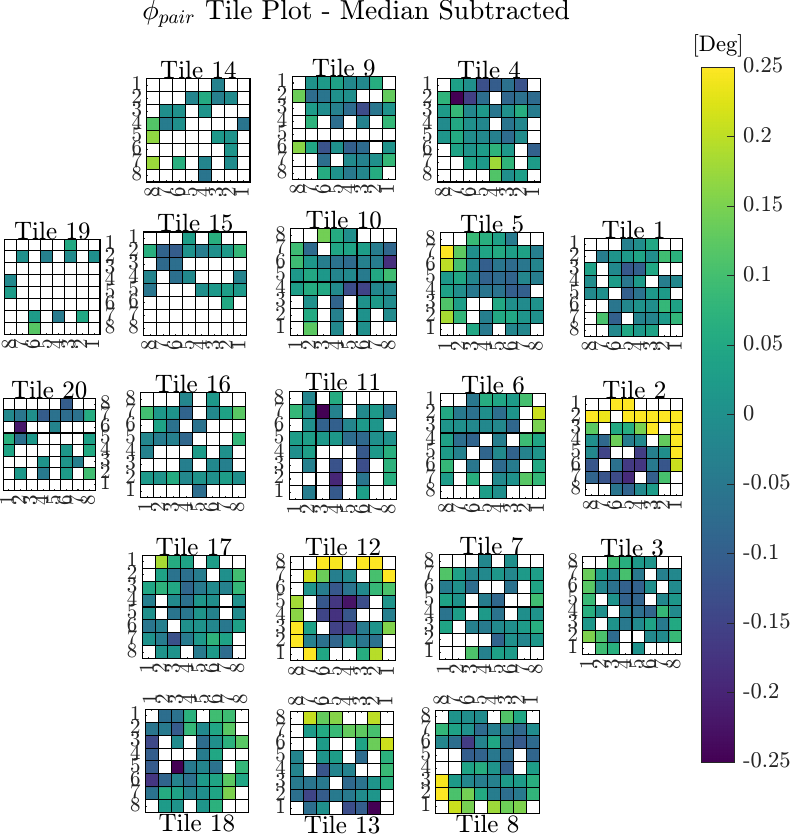}
    \vspace{-5pt}
    \caption{Left, 2021\textendash22 season measurements of the B3 focal plane. Right, per-tile median-subtracted version of left. Figures from \cite{spie_proc}.}
    \label{fig:b3fpu}
    \vspace{-15pt}
\end{figure}

In the 2021\textendash22 season, we executed a successful calibration campaign using the RPS \cite{bkxviii}. 
Figure \ref{fig:b3fpu} shows the results. 
We measure $\sim0.28^\circ$ variation in `clocking' between individual detector modules, which are physically separate and each independently bolted to the focal plane. 
Subtracting the per-tile medians reveals $\sim 0.065^\circ$ variations within tiles (likely from the fabrication process).
These angle variations are in good agreement with our CMB-derived detector pointing model.
However, the measurements were limited by a systematic uncertainty due to variations in polarization across the RPS beam, introducing an unknown offset in the polarization angle. 

\subsection{2024\textendash25 RPS tests at Pole}
\begin{figure}
    \centering
    \includegraphics[width=0.8\linewidth]{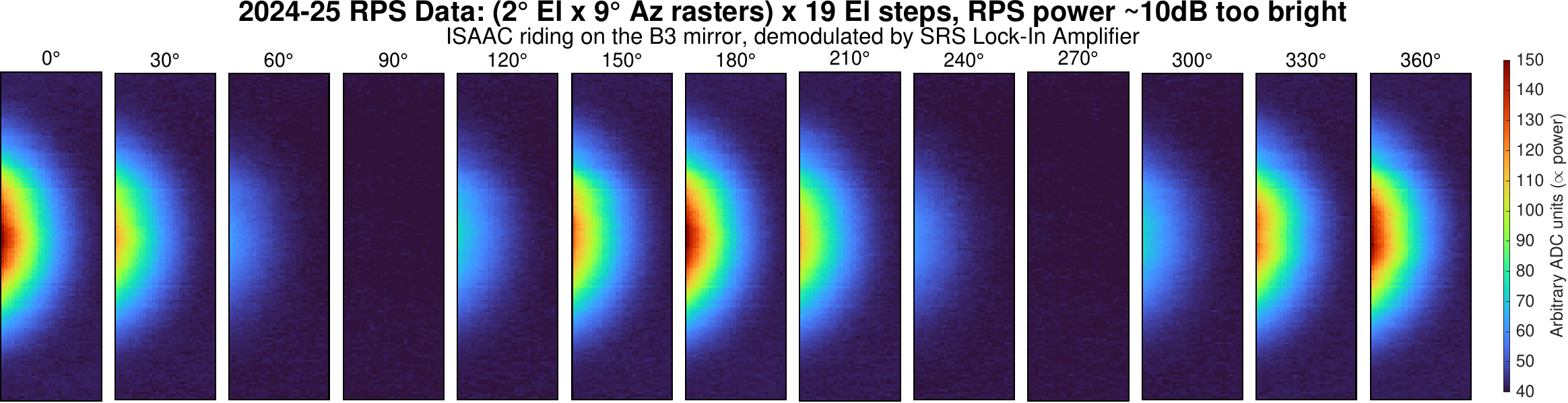}
    \vspace{-5pt}
    \caption{Successful in-situ observation of the RPS with the ISAAC at Pole, at a range of 200m. }
    \label{fig:isaac_data}
    \vspace{-15pt}
\end{figure}

The ISAAC has been used in-situ for the first time, and will provide an independent polarization angle verification for future RPS campaigns. As shown in Figure \ref{fig:isaac_data}, the structure of the modcurve is visible at high SNR, despite the beam being offset to couple with the center of the mirror and thus the B3 receiver.

\section{Mitigation of RF systematics}
In \cite{bkxviii}, we found that our error budget was dominated by spatially varying polarization in the RPS beam. 
Departures from uniform polarization convert small and unavoidable pointing errors (e.g., mast sway) into significant polarization errors.
In the current RPS configuration, this effect is strong enough ($\sim\pm0.3^\circ$) to dominate our error budget.

We have demonstrated that this pointing-dependent polarization is due to power incident on the conductive aperture of the wire grid. This power is partially diffracted into unwanted cross-polarization, creating a non-zero, non-uniform U beam, as shown in both simulations and measurements in the lab.

\begin{figure}
    \centering
    \includegraphics[width=0.7\linewidth]{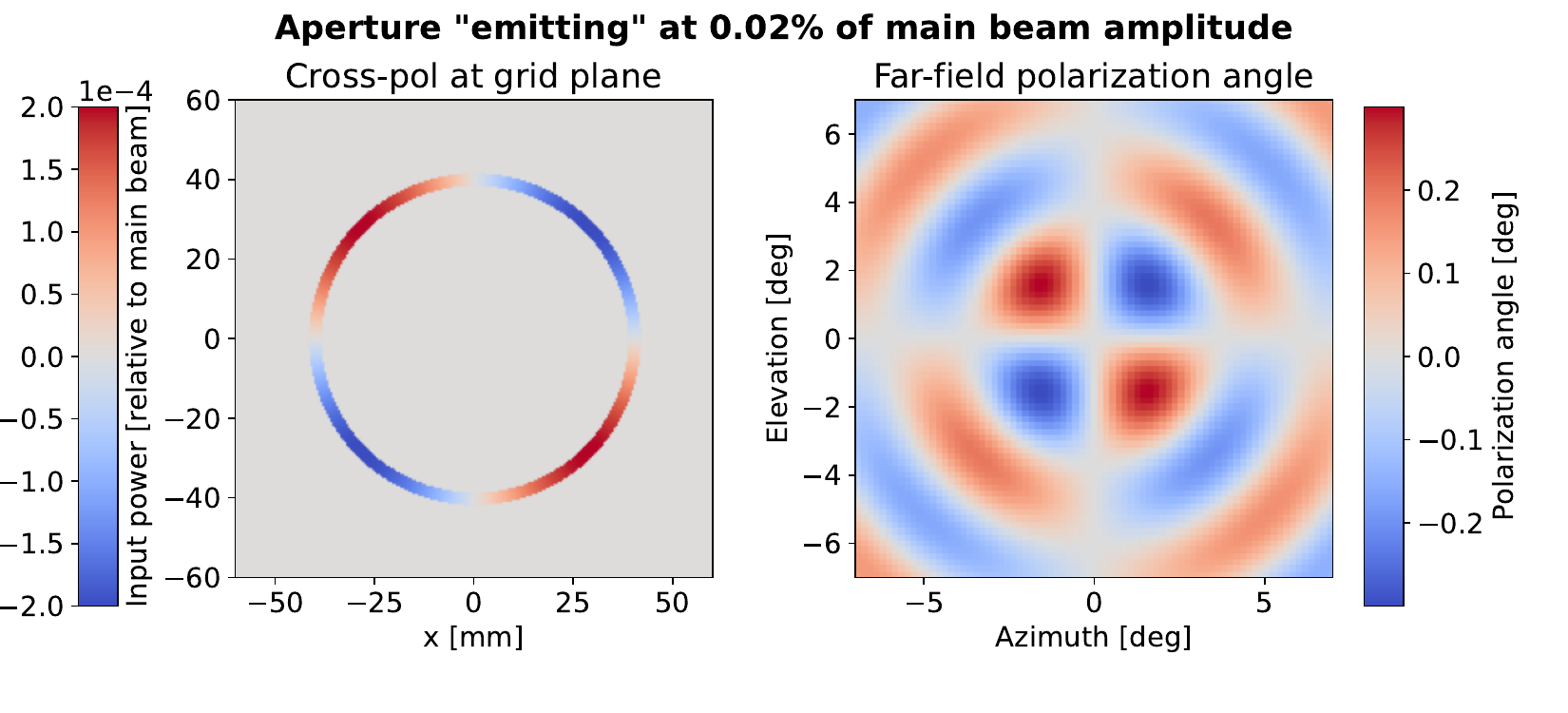}
    \vspace{-10pt}
    \caption{\textbf{Left}, simplified model of cross-pol from diffraction off the grid's aperture. \textbf{Right}, Fourier transform of left, showing how the illuminated ring generates polarization errors in the far field.}
    \label{fig:pol_model}
    \vspace{-15pt}
\end{figure}

\begin{figure}
    \centering
    \includegraphics[width=0.9\linewidth]{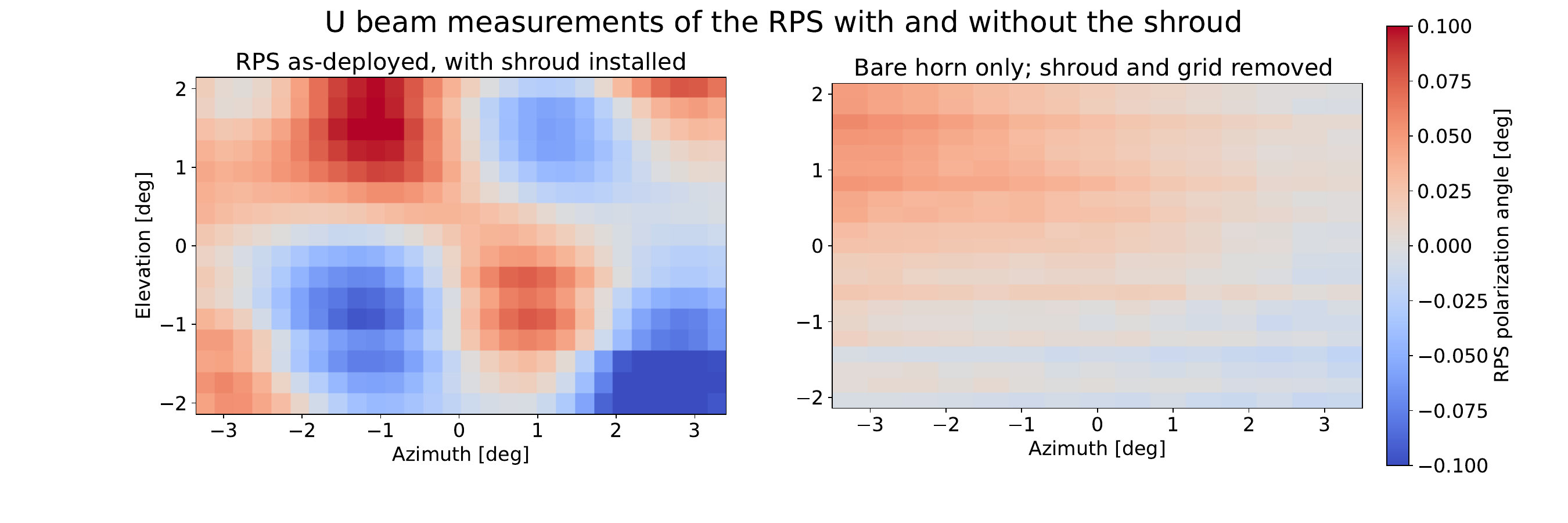}
    \vspace{-10pt}
    \caption{Measured Stokes U beam of the RPS. \textbf{Left}, the full RPS system showing the cross-pol morphology predicted by the model. \textbf{Right}, same setup as left but without the shroud or wire grid.}
    \label{fig:pol_data}
    \vspace{-15pt}
\end{figure}

Work is ongoing to mitigate this issue by replacing the horn with a low-sidelobe corrugated horn and installing stops inside the shroud to better block lines of sight between the horn and the wire grid aperture.

\section{Conclusions}
We have now demonstrated in-situ operations of both the Rotating Polarized Source calibrator and the In-Situ Absolute Angle Calibrator. 
In \cite{bkxviii}, we took useful measurements of BICEP3, but they are limited by systematic uncertainty.
We have now understood the source of the systematic uncertainty (diffraction off the wire grid), and are working to mitigate it.
If the systematic uncertainty can be mitigated to better than 0.02$^\circ$, then we can measure uniform cosmic birefringence to our target measurement accuracy of $\mathcal{O}(0.05^\circ)$.

%
% BibTeX or Biber users please use (the style is already called in the class, ensure that the "woc.bst" style is in your local directory)

%\bibliography{poster}
%\nocite{*}
% To cite an article \cite{bleem:2024}.
% \begin{thebibliography}{}
% %
% % and use \bibitem to create references.
% %
% %Please, keep the bibliography section rather compact. First author et al. is enough for most of the papers. No title.
% %Please follow the style : Journal \textbf{Vol}, first page (Year)

\end{document}

%% file: author_list_mmuniverse_2025.tex
% To use this file: simply \input{author_list_<paper_id>.tex} in your main TeX file where you want the author list to go!
% generated for paper: mmUniverse 2025
% 2025-09-26 16:11:14

% hand-edited into the format of this doc
\author{
\lastname{A.~R.~Polish}\inst{5,15}\fnsep\thanks{e-mail: {apolish@g.harvard.edu}} \and
\lastname{P.~A.~R.~Ade}\inst{1} \and
\lastname{Z.~Ahmed}\inst{2,3} \and
\lastname{M.~Amiri}\inst{4} \and
\lastname{D.~Barkats}\inst{5} \and
\lastname{R.~Basu~Thakur}\inst{6} \and
\lastname{C.~A.~Bischoff}\inst{7} \and
\lastname{D.~Beck}\inst{8} \and
\lastname{J.~J.~Bock}\inst{6,9} \and
\lastname{H.~Boenish}\inst{5} \and
\lastname{V.~Buza}\inst{10} \and
\lastname{B.~Cantrall}\inst{8,2} \and
\lastname{J.~R.~Cheshire~IV}\inst{6} \and
\lastname{J.~Connors}\inst{11} \and
\lastname{J.~Cornelison}\inst{12} \and
\lastname{M.~Crumrine}\inst{13} \and
\lastname{A.~J.~Cukierman}\inst{6} \and
\lastname{E.~Denison}\inst{11} \and
\lastname{L.~Duband}\inst{14} \and
\lastname{M.~Echter}\inst{5} \and
\lastname{M.~Eiben}\inst{5} \and
\lastname{B.~D.~Elwood}\inst{5,15} \and
\lastname{S.~Fatigoni}\inst{6} \and
\lastname{J.~P.~Filippini}\inst{16,17} \and
\lastname{A.~Fortes}\inst{8} \and
\lastname{M.~Gao}\inst{6} \and
\lastname{C.~Giannakopoulos}\inst{7} \and
\lastname{N.~Goeckner-Wald}\inst{8} \and
\lastname{D.~C.~Goldfinger}\inst{8} \and
\lastname{S.~Gratton}\inst{18,19} \and
\lastname{J.~A.~Grayson}\inst{8} \and
\lastname{A.~Greathouse}\inst{6} \and
\lastname{P.~K.~Grimes}\inst{5} \and
\lastname{G.~Hall}\inst{20,8} \and
\lastname{G.~Halal}\inst{8} \and
\lastname{M.~Halpern}\inst{4} \and
\lastname{E.~Hand}\inst{7} \and
\lastname{S.~A.~Harrison}\inst{5} \and
\lastname{S.~Henderson}\inst{2,3} \and
\lastname{T.~D.~Hoang}\inst{13} \and
\lastname{J.~Hubmayr}\inst{11} \and
\lastname{H.~Hui}\inst{6} \and
\lastname{K.~D.~Irwin}\inst{8} \and
\lastname{J.~H.~Kang}\inst{6} \and
\lastname{K.~S.~Karkare}\inst{21} \and
\lastname{S.~Kefeli}\inst{6} \and
\lastname{J.~M.~Kovac}\inst{5,15} \and
\lastname{C.~Kuo}\inst{8} \and
\lastname{K.~Lasko}\inst{13,20} \and
\lastname{K.~K.~Lau}\inst{6} \and
\lastname{M.~Lautzenhiser}\inst{7} \and
\lastname{A.~Lennox}\inst{17} \and
\lastname{T.~Liu}\inst{8} \and
\lastname{S.~Mackey}\inst{10} \and
\lastname{N.~Maher}\inst{13} \and
\lastname{K.~G.~Megerian}\inst{9} \and
\lastname{L.~Minutolo}\inst{6} \and
\lastname{L.~Moncelsi}\inst{6} \and
\lastname{Y.~Nakato}\inst{8} \and
\lastname{H.~T.~Nguyen}\inst{6,9} \and
\lastname{R.~O’Brient}\inst{6,9} \and
\lastname{S.~N.~Paine}\inst{5} \and
\lastname{A.~Patel}\inst{6} \and
\lastname{M.~A.~Petroff}\inst{5} \and
\lastname{T.~Prouve}\inst{14} \and
\lastname{C.~Pryke}\inst{13} \and
\lastname{C.~D.~Reintsema}\inst{11} \and
\lastname{T.~Romand}\inst{6} \and
\lastname{M.~Salatino}\inst{8} \and
\lastname{A.~Schillaci}\inst{6} \and
\lastname{B.~Schmitt}\inst{5} \and
\lastname{B.~Singari}\inst{13,20} \and
\lastname{A.~Soliman}\inst{6,9} \and
\lastname{T.~St.~Germaine}\inst{5} \and
\lastname{A.~Steiger}\inst{6} \and
\lastname{B.~Steinbach}\inst{6} \and
\lastname{R.~Sudiwala}\inst{1} \and
\lastname{K.~L.~Thompson}\inst{2,8} \and
\lastname{C.~Tsai}\inst{5} \and
\lastname{C.~Tucker}\inst{1} \and
\lastname{A.~D.~Turner}\inst{9} \and
\lastname{C.~Verg\`{e}s}\inst{5} \and
\lastname{A.~G.~Vieregg}\inst{10} \and
\lastname{A.~Wandui}\inst{6} \and
\lastname{A.~C.~Weber}\inst{9} \and
\lastname{J.~Willmert}\inst{13} \and
\lastname{W.~L.~K.~Wu}\inst{2,3} \and
\lastname{H.~Yang}\inst{8} \and
\lastname{C.~Yu}\inst{10} \and
\lastname{L.~Zheng}\inst{5} \and
\lastname{C.~Zhang}\inst{6} \and
\lastname{S.~Zhang}\inst{6}
}

\institute{
    School of Physics and Astronomy, Cardiff University, Cardiff, CF24 3AA, United Kingdom
    \and Kavli Institute for Particle Astrophysics \& Cosmology, Stanford University, Stanford, CA 94305, USA
    \and SLAC National Accelerator Laboratory, Menlo Park, CA 94025, USA
    \and Department of Physics and Astronomy, University of British Columbia, Vancouver, British Columbia, V6T 1Z1, Canada
    \and Center for Astrophysics | Harvard \& Smithsonian, Cambridge, MA 01238, USA
    \and Department of Physics, California Institute of Technology, Pasadena, CA 91125, USA
    \and Department of Physics, University of Cincinnati, Cincinnati, OH 45221, USA
    \and Department of Physics, Stanford University, Stanford, California 94305, USA
    \and Jet Propulsion Laboratory, California Institute of Technology, Pasadena, CA 91109, USA
    \and Kavli Institute for Cosmological Physics, University of Chicago, Chicago, IL 60637, USA
    \and National Institute of Standards and Technology, Boulder, CO 80305, USA
    \and High-Energy Physics Division, Argonne National Laboratory, Lemont, IL, 60439, USA
    \and School of Physics and Astronomy, University of Minnesota, Minneapolis, MN 55455, USA
    \and Service des Basses Temp\'eratures, Commissariat \`a l'\'Energie Atomique, 38054 Grenoble, France
    \and Department of Physics, Harvard University, Cambridge, MA 02138, USA
    \and Department of Physics, University of Illinois at Urbana-Champaign, Urbana, Illinois 61801, USA
    \and Department of Astronomy, University of Illinois at Urbana-Champaign, Urbana, Illinois 61801, USA
    \and Centre for Theoretical Cosmology, DAMTP, University of Cambridge, Cambridge CB3 0WA, UK
    \and Kavli Institute for Cosmology Cambridge, Cambridge CB3 0HA, UK
    \and Minnesota Institute for Astrophysics, University of Minnesota, Minneapolis, MN 55455, USA
    \and Department of Physics, Boston University, Boston, MA 02215, USA
}